\documentclass{cpbtex}

\usepackage{graphicx}

\begin{document}

\title{Higher-order topology in  twisted multilayer systems: a review}

\author{Chunbo Hua$^{1,2}$, \ Dong-Hui Xu$^{3,4}$\thanks{Corresponding author. E-mail:donghuixu@cqu.edu.cn}\\
$^{1}${Key Laboratory of Artificial Micro- and Nanostructures of Ministry of Education}\\
{and School of Physics and Technology, Wuhan University, Wuhan 430072, China}\\  
$^{2}${School of Electronic and Information Engineering,}\\
{Hubei University of Science and Technology, Xianning 437100, China} \\
$^{3}${Institute for Structure and Function and Department of Physics}\\
{and Chongqing Key Laboratory for Strongly Coupled Physics,}\\
{Chongqing University, Chongqing 400044, China}\\ 
$^{4}${Center of Quantum Materials and Devices, Chongqing University, Chongqing 400044, China}}   


\date{\today}
\maketitle

\begin{abstract}
In recent years, there has been a surge of interest in higher-order topological phases (HOTPs) across various disciplines within the field of physics. These unique phases are characterized by their ability to harbor topological protected boundary states at lower-dimensional boundaries, a distinguishing feature that sets them apart from conventional topological phases and is attributed to the higher-order bulk-boundary correspondence. Two-dimensional (2D) twisted systems offer an optimal platform for investigating HOTPs, owing to their strong controllability and experimental feasibility. Here, we provide a comprehensive overview of the latest research advancements on HOTPs in 2D twisted multilayer systems. We will mainly review the HOTPs in electronic, magnonic, acoustic, photonic and mechanical twisted systems, and finally provide a perspective of this topic.
\end{abstract}



\section{Introduction}

In the past two decades, the field of condensed matter physics has witnessed a significant surge in research interest and activity surrounding the topological phase of matter. In recent years, a new class of topological phases known as higher-order topological phase (HOTPs) has been discovered\ucite{Benalcazar2017Science,Benalcazar2017PRB,Langbehn2017PRL,Song2017PRL,Schindler2018SA,Yan2019Wlxb,Xie2021NRP,Yang2024JPCM,Wieder2022NRM}. The HOTPs demonstrate $d$-dimensional $n$th-order topological phases that accommodate ($d-n$)-dimensional boundary states with codimension $n>1$ due to the higher-order bulk-boundary correspondence, which is distinct from conventional topological phases in $d$ dimensions hosting ($d-1$)-dimensional boundary states as per the conventional bulk-boundary correspondence. For instance, a two-dimensional (2D) second-order topological phase exhibits gapped 1D edge states and 0D corner states within the energy gap, as shown in Fig.~1(a). A 3D second-order topological phase displays gapped 2D surface states and gapless 1D hinge states, while a 3D third-order topological phase features gapped surface and hinge states, as well as in gap corner states (Fig.~1(b)). In 2017, Benalcazar $et~al.$\ucite{Benalcazar2017Science,Benalcazar2017PRB} proposed a framework for multipole topological insulators protected by mirror symmetries, now known as the Benalcazar-Bernevig-Hughes (BBH) model. This model showcases in-gap corner states and quantized multipole moments as the distinctive hallmark of higher-order bulk-boundary correspondence for higher-order topology. Subsequently, these has been a significant advancement in both theoretical and experimental research within this field\ucite{Garcia2018Nature,Peterson2018Nature,Schindler2018NP,Imhof2018NP,Xu2019PRLHOTI,Yue2019NP,Xue2019NM,Ni2019NM,Sheng2019PRL,Ren2020PRL,Zhang2020PRL,Chen2021PRL,
Ning2022PRB,ChenR2023PRB,XuXX2024CPB,Zhan2024NL,ChneH2023CPB,YanZB2018PRL,YanZB2019PRL,YanZB2019PRB,ZhuD2023PRB,ZengQB2020PRB,YangYB2020PRR,Tao2020NJP,YangYB2021PRB,PengY2022OL,ShiA2024PRB,ZhaoPL2021PRL,
LiuT2018PRB,LiuT2019PRL,LiuT2021PRL,LuC2023PRB,QiY2020PRL,WuJ2020PRB,YangY2021PRL,WeiQ2021PRL,WuJ2022PRB,WangZ2024CP,YeL2024PRL,HuangX2022SB,QiY2022APL,QiY2024PRB,PuZ2023PRL,MaQ2024PRL,
LuoXJ2023PRB,LiuF2023PRB,LuoXJ2023PRB2,HuYS2021PRB,LiuZ2022PRB,LiuZR2021PRB,LuC2023PRB2,HuY2024PRB,ChenR2024PRB,PengT2024PRB,WuYJ2022PRA,LiCA2022FP,ChneXT2023CPB,
LiJ2023APL,Guo2023MTP,Miao2023PRB,Scammell2022PRB,Habel2024PRB}, leading to the study of numerous systems featuring higher-order topology. Furthermore, non-periodic systems are also capable of discovering HOTPs in addition to crystalline systems, such as quasicrystalline systems\ucite{Varjas2019PRL,Chen2020PRL,Hua2020PRB,Huang2021NL,Peng2021PRB,Wang2022PRL,Xiong2022PRAp,Mao2024PRB,Ouyang2024PRB,Lv2021CP,Spurrier2020PRR,Traverso2022Symmetry,Traverso2022PRB,Shi2024Laser}, amorphous systems\ucite{Agarwala2020PRR,Wang2021PRLAmorphous,Peng2022PRB,Tao2023Post}, fractal systems\ucite{Manna2022PRB,Zheng2022SB,ChenH2023CPB} and hyperbolic systems\ucite{Tao2023PRB,Liu2023PRB}.

\begin{center}
\includegraphics[width=10cm]{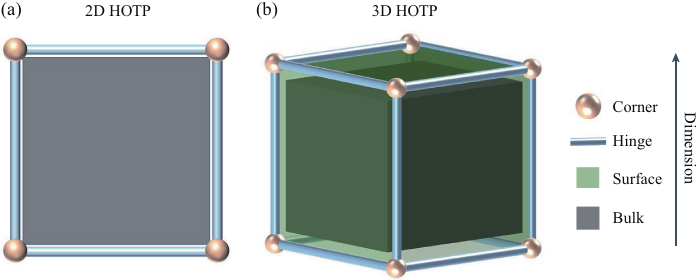}\\[5pt]  
\parbox[c]{15.0cm}{\footnotesize{\bf Fig.~1.} The categorization of HOTPs characterized by multidimensional topological boundary states is presented\ucite{Xie2021NRP}. (a) For 2D HOTP depicted by grey areas, and 0D topological corner states represented as pink spheres are shown. (b) For 3D HOTP, there are 1D topological hinge states and 0D topological corner states.}
\end{center}

On the other hand, 2D twisted van der Waals materials have emerged as a versatile platform for investigating exotic and elusive states of matter, subsequent to the discovery of unconventional superconductivity\ucite{Cao2018Nature1} and the Mott insulator\ucite{Cao2018Nature2} in twisted bilayer graphene with magic angles\ucite{Bistritzer2011PNAS}. The engineering of twisted moir\'{e} superlattices introduces a new freedom for manipulating the physics of 2D van der Waals materials, with the twist playing a pivotal role in modulating the quantum phases of these systems. For instance, Fig.~2 depicts a schematic representation of twisted moir\'{e} superlattices with the large commensurate angle $21.78^\circ$. A range of novel quantum phases have been predicted or experimentally observed in 2D twisted van der Waals materials, including quantum anomalous Hall phases\ucite{Serlin2020Science}, quantum spin Hall phases\ucite{Wu2019PRL}, as well as higher-order topological phases\ucite{Park2019PRL,Liu2021PRL,Park2021Carbon,Liu2022PRB,Lim2023njpCM,Chew2023PRB,Qian2023PRB,Spurrier2020PRR,Traverso2022Symmetry,Traverso2022PRB,Shi2024Laser,Hua2023PRB,Wu2022PRAp,Oudich2021PRB,Yi2022Light,Danawe2021PRB}.

The unexpected correlation between higher-order topology and 2D twisted van der Waals materials was first observed in twisted bilayer graphene (TBG) by Park $et~al.$\ucite{Park2019PRL} in 2019. In Ref.~\cite{Liu2021PRL}, the authors predicted the emergence of a 2D second-order topological insulator in twisted bilayer graphene and twisted bilayer boron nitride, with both zero and full filling gaps. A proposed instanton approach argues that the tunneling of electrons between the higher-order topological corner states of the higher-order topological generally leads to gate-tunable oscillation in the energy spectra\ucite{Park2021Carbon}. A second-order topological insulator has also been identified in The twisted bilayer $\alpha$-graphyne at large twisting angle\ucite{Liu2022PRB}. Moreover, there has been a growing research interest in the exploration of higher-order topological superconductors\ucite{Chew2023PRB} and higher-order topological semimetals\ucite{Qian2023PRB} within electronic twisted van der Waals systems.
Furthermore, the study of HOTPs in 2D twisted van der Waals has also been extended to bosonic magnonic systems\ucite{Hua2023PRB}, as well as acoustic\ucite{Wu2022PRAp}, photonic\ucite{Oudich2021PRB,Yi2022Light}, and mechanical systems\ucite{Danawe2021PRB}.

\begin{center}
\includegraphics[width=10cm]{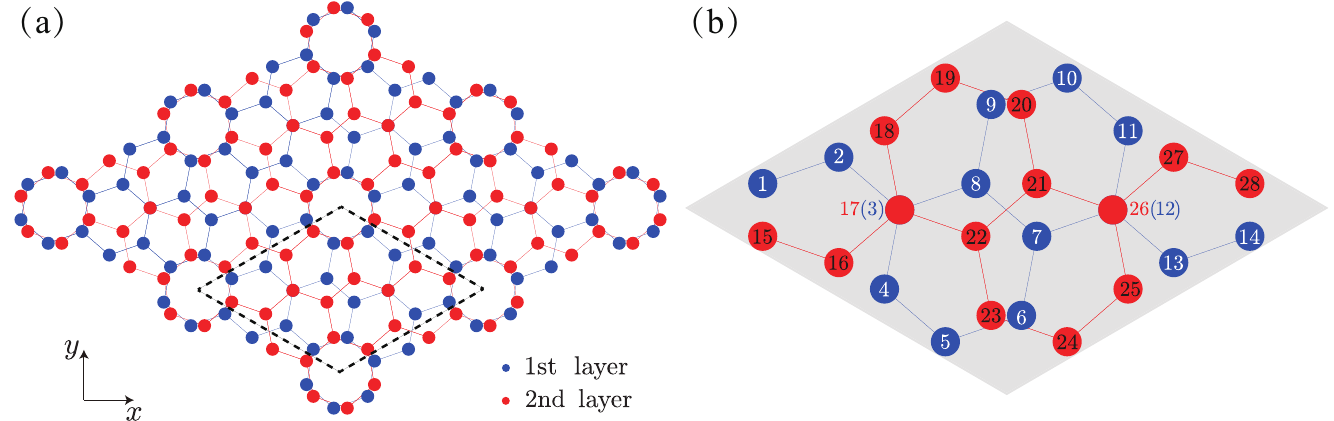}\\[5pt]  
\parbox[c]{15.0cm}{\footnotesize{\bf Fig.~2.} (a) Schematic illustration of the twisted moir\'{e} superlattices with the large commensurate angle $21.78^\circ$. The rhombus-shaped region outlined by thick black dashed lines represents a single moir\'{e} unit cell. (b) Schematic depiction of a unit cell for the lattice, comprising 28 lattice sites, each labeled with a number. The first layer lattice is denoted by blue dots, while the second layer lattice is represented by red dots. The blue (red) lines indicate the intralayer nearest-neighbor bonds within the first (second) layer.\ucite{Hua2023PRB}}
\end{center}


In this review, we showcase the achievements in the emerging field of HOTPs in 2D twisted van der Waals materials and 2D twisted metamaterials. Section 2 provides a comprehensive overview of HOTPs in 2D electronic twisted van der Waals materials. Subsequently, section 3 delves into an examination of studies on HOTPs in magnonic twisted systems. In section 4, we survey research on HOTPs in other classical twisted systems, encompassing acoustic, photonic and mechanical systems. Finally, our review culminates with a summary and outlook presented in section 5.

\section{Higher-order topology in electronic twisted systems}

In this section, we will review HOTPs in 2D electronic twisted van der Waals materials, including higher-order topological insulator phases\ucite{Park2019PRL,Liu2021PRL,Park2021Carbon,Liu2022PRB,Lim2023njpCM}, higher-order topological superconductor phases\ucite{Chew2023PRB}, higher-order topological semimetal phases\ucite{Qian2023PRB}, and quasicrystalline higher-order topological insulator phases\ucite{Spurrier2020PRR,Traverso2022Symmetry,Traverso2022PRB,Shi2024Laser}.

In 2019, Park $et~al.$\ucite{Park2019PRL} proposed the TBG with large commensurate angle of $21.78^\circ$ as higher-order topological insulators, hosting topological corner states. The atomic configuration of the TBG can be easily constructed by twisting AA-stacked bilayer graphene relative to the collinear axis at the hexagonal center. Furthermore, the system maintains rotational symmetries of $C_{2x}$, $C_{2y}$ and $C_{6z}$ about the in-plane $x$, $y$ and out-of-plane $z$ axes, respectively. Figure~3(a) illustrates electronic energy band structures of the TBG with twist angle $21.78^\circ$ using first-principles calculations. The results reveal that the band structure features a narrow gap of approximately $9$ meV, with a global direct band gap appearing between conduction and valence bands, while the minimum value occurs in proximity to point K as shown in the right panel of Fig. 3(a). The existence of a band gap due to the broken $U(1)_v$ symmetry at large twist angles, while the increased intervalley coupling results in the opening of a gap between these Dirac points.  The higher-order topology at half filling is directly calculated by the mirror-winding number and the second Stiefel-Whitney number. In addition, the higher-order topology at half filling, where the gap is present at $K$ point, in the TBG at all commensurate angles was discussed.

\begin{center}
\includegraphics[width=10cm]{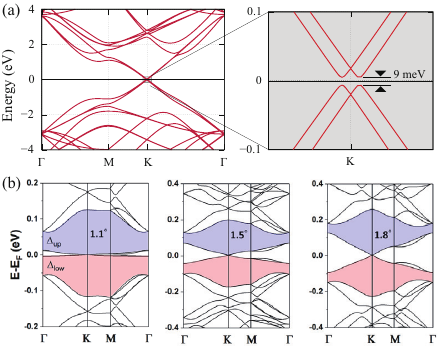}\\[5pt]  
\parbox[c]{15.0cm}{\footnotesize{\bf Fig.~3.}   Higher-order topology of TBG with different commensurate twist angles. (a) Electronic energy band structures of TBG with large commensurate angle $21.78^\circ$\ucite{Park2019PRL}. The right panel shows the magnified view of the region near K indicated by a gray box in the left panel. (b) Electronic energy band structures of TBG with small large commensurate angle $1.1^\circ$, $1.5^\circ$ and $1.8^\circ$\ucite{Liu2021PRL}. The blue- and red-color region denotes the upper and lower gap for isolating the low-energy bands from other high-energy bands.}
\end{center}

However, the nontrivial gap at low-energy Dirac bands in TBG with twist angle $21.78^\circ$ is extremely small (9 meV), making it difficult to detect in-gap topological corner states in experiments. Liu $et~al.$\ucite{Liu2021PRL} have theoretically predicted the existence of a 2D second-order topological insulator in twisted bilayer graphene and twisted bilayer boron nitride, with both zero and full filling gaps. And three distinct characteristics of the 2D second-order topological insulator were identified: a nonzero bulk topological index, gapped topological edge states, and in-gap topological corner states. In addition, large high-energy band gaps were found, where the largest gap goes up to $60$ meV for TBG with twist angle $0.9^\circ$-$2.5^\circ$ and $102$ meV for twisted bilayer boron nitride with twist angle $0.8^\circ$-$5.1^\circ$. The typical band structures of TBG with different commensurate twist angles are shown in Figs. 3(b).

Additionally, the phenomenon of instanton tunneling between the corner states in the HOTPs of TBG has been investigated\ucite{Park2021Carbon}. And, it has been observed that twisted bilayer $\alpha$-graphyne at a twist angle of $21.78^\circ$ also exhibits characteristics of a higher-order topological insulator, with a significantly larger bulk band gap of $12$ meV compared to TBG\ucite{Liu2022PRB}. Furthermore, The occurrence of HOTPs in TBG is notable due to the absence of explicit protecting symmetries, which can be attributed to the self-similarity of Hofstadter butterflies as replicas of original HOTPs\ucite{Lim2023njpCM}.

The kagome lattice, a network of triangles connected by their corners, possesses a unique electronic structure characterized by Dirac cones, van Hove singularities, and flat bands. This distinctive arrangement makes it an ideal platform for investigating the intricate interplay of frustrated geometry, electron correlations, and topology. Recent investigations have unveiled a series of kagome materials exhibiting exotic quantum phenomena. These materials demonstrate a remarkable confluence of intertwined charge order and superconductivity, modulated magnetism, and non-trivial topological states, establishing them as a fertile ground for exploring emergent quantum behavior\ucite{Yin2022Nature, Neupert2022NP,Wang2023NRP,Wilson2024NRM}. The combination of kagome and moir\'{e} physics leads to fascinating phenomena. For example, in kagome twisted bilayers, two sets of flat bands appear at distinct energies\ucite{Crasto2019PRB}. Furthermore, these bilayers can exhibit higher-order topological insulator at large commensurate angles\ucite{Wan2024arxiv}. The higher-order topological insulator exhibits corner states and is independent protected by either rational symmetry $C_{3z}$ or mirror symmetry.

Moreover, the emergence of higher-order topological superconductor phases has been observed through the proximitization of TBG with spin-singlet (or spin-triplet) superconductivity\ucite{Chew2023PRB}. Multiple instances of $C_{2z}T$-protected Majorana Kramers pairs manifest  at corners on pairing domain walls. This phenomenon in the pairing behavior of TBG is a consequence of the anomaly ensured by approximate particle-hole symmetry and $C_{2z}T$. Additionally, it is proposed that alternating-twist multilayer graphene, which can be regarded as 3D TBG, serves as a platform for stable higher-order topological Dirac semimetals with higer-order hinge Fermi arcs\ucite{Qian2023PRB}.

\begin{center}
\includegraphics[width=15cm]{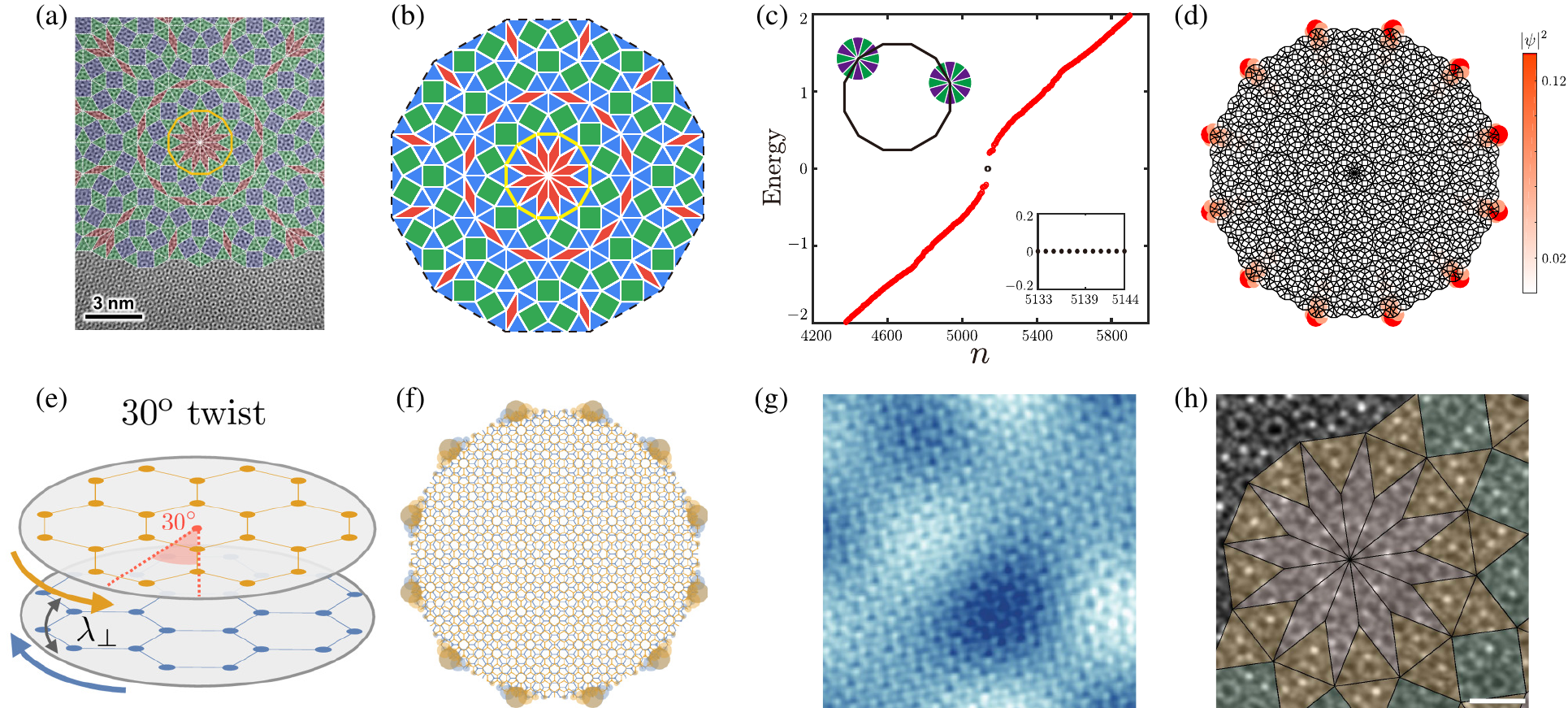}\\[5pt]  
\parbox[c]{15.0cm}{\footnotesize{\bf Fig.~4.} (a) TEM image of graphene quasicrystal mapped with 12-fold Stampfli-inflation tiling\ucite{Ahn2018Science}. (b) Schematic illustrations of the Stampfli-tiling quasicrystal dodecagon\ucite{Hua2020PRB}.
(c) Energy spectrum of the quasicrystalline HOTP versus the eigenvalue index $n$. The inset at the top displays the color circle representing the effective edge mass, with green and violet regions denoting two opposite orientations of the edge with respect to the sign of the effective edge mass. The inset at lower right provides an enlarged view of 12 zero-energy modes, indicated by black dots.\ucite{Hua2020PRB} (d) The probability density of zero-energy modes in (c) is depicted using a color map to represent its values. (e) Schematic illustration of the Kane-Mele model amounts to stacking two Haldane layers with opposite Chern numbers and $30^\circ$ twist.\ucite{Spurrier2020PRR} (f) The probability density of corner-localized modes of (e). 
(g) Typical STM topography for WSe$_{2}$ incommensurate quasicrystal\ucite{Li2024Nature}. (h) STEM-ADF image of $30^\circ$ twisted-bilayer quasicrystal MoS$_{2}$ mapped with 12-fold Stampfli-inflation tiling\ucite{Tsang2024Science}.}
\end{center}

The concept of HOTPs has been expanded to quasicrystalline systems\ucite{Chen2020PRL}, where conventional classification of topological phases with crystal-compatible symmetries are inadequate. A unique higher-order topological insulator protected by an eightfold rotational symmetry, which is not present in crystalline materials, has been discovered in Ammann-Beenker tiling octagonal quasicrystals\ucite{Chen2020PRL}. Recently, the extrinsic dodecagonal quasicrystalline lattice was achieved in the twisted bilayer graphene precisely rotated at an angle of $30^\circ$\ucite{Ahn2018Science,Yao2018PNAS}, as depicted in Fig 4(a). Then, the authors in~\cite{Hua2020PRB} have demonstrated the existence of a quasicrystalline higher-order topological insulator in an intrinsic dodecagonal quasicrystal (Fig 4(b)). The quasicrystalline HOTP exhibits twelve in-gap zero-energy modes symmetrically distributed at the corners of a quasicrystal dodecagon, as depicted in Figs. 4(c) and (d).  Additionally, These zero-energy corner modes are protected by a combination of the twelvefold rotational symmetry and mirror symmetry, as well as particle-hole symmetry, which lacks a crystalline counterpart.  At the same time, Spurrier $et~al.$\ucite{Spurrier2020PRR} proposed the existence of HOTPs in an extrinsic dodecagonal quasicrystalline lattice. Their model consists of two stacked Haldane models with oppositely propagating edge modes, similar to the Kane-Mele model but with a $30^\circ$ twist, as shown in Fig. 4(e). The localized modes observed at corners in Fig. 4(f), which are characteristic of a quasicrystalline HOTP, do not stem from conventional mass inversions but rather arise from the fractional mass kinks. Very recently, there has been extensive research on the incommensurate extrinsic dodecagon quasicrystals at $30^\circ$, including the experimental observation of WSe$_{2}$\ucite{Li2024Nature} and MoS$_{2}$\ucite{Tsang2024Science} incommensurate quasicrystals. Hence, it is anticipated that experimental discoveries of quasicrystalline higher-order topological phases in twisted bilayer systems will be forthcoming.

\section{Higher-order topology in magnonic twisted systems}

Recently, the concept of higher-order topology has been extended to magnonic systems as well\ucite{Sil2020JPCM,Hirosawa2020PRL,Mook2021PRB,Li2020PRB,Li2021PRB,Park2021PRB,Li2022PRB,Bhowmik2024PRB}. For instance, a second-order topological magnon insulator (SOTMI) with magnon corner states is realized in a ferromagnetic (FM) Heisenberg model on a 2D breathing kagome lattice\ucite{Sil2020JPCM}, a magnonic quadrupole topological insulator hosting magnon corner states can appear in 2D antiskyrmion crystals\ucite{Hirosawa2020PRL}, and a SOTMI with 1D chiral hinge magnons is predicted to be realized in 3D stacked honeycomb magnets\ucite{Mook2021PRB}. Meanwhile, researchers have shifted their focus to twisted bilayer honeycomb magnets (TBHMs) analogous to twisted bilayer graphene, and have uncovered a diverse range of magnetic properties resulting from moir\'{e} patterns\ucite{Ahn2024PQE,ChengR2020PRB,Wang2020PRL,Wang2023PRX,Bostrom2020Post,Kim2023NL,LiuJ2020PRB,Zhu2021CPB,Ghader2020SR,Hejazi2020PNAS}.

In this following, we present the realization of a SOTMI in a TBHM at the large commensurate angle $21.78^{\circ}$\ucite{Hua2023PRB}, without necessitating the presence of Dzyaloshinskii-Moriya interaction. The TBHM is specifically engineered by twisting AA-stacked bilayer honeycomb magnets to achieve the commensurate angle $21.78^{\circ}$, with spins localized at the hexagon vertices. The schematic illustration of a unit cell for the TBHMs lattice is shown in Fig. 2(b). A minimal spin model is utilized to describe the TBHM with collinear order, consisting of two honeycomb FM layers with nearest-neighbor intralayer exchange interaction coupled by FM interlayer exchange coupling or antiferromagnetic (AFM) interlayer exchange coupling. The spin Hamiltonian is formulated for the twisted bilayer honeycomb lattice, which reads
\begin{equation}
H=-J\sum_{\left\langle i,j\right\rangle ,l}\mathbf{S}_{i,l}\cdot \mathbf{S}%
_{j,l}-J_{\bot }\sum_{\left\langle i,j\right\rangle }\mathbf{S}_{i,2}\cdot
\mathbf{S}_{j,1},
\end{equation}
where the first and second terms represent the nearest-neighbor intralayer and nearest-neighbor interlayer Heisenberg interactions, respectively. $\mathbf{S}_{i,l}=(S^{x}_{i,l},S^{y}_{i,l},S^{z}_{i,l})$ is the spin vector operator at site $i$ on layer $l=1,2$, and the summation runs over nearest-neighbor sites $\left\langle i,j\right\rangle$. $J>0$ denotes the FM intralayer interaction, and $J_{\bot }$ is a tunable parameter in TBHMs, which is positive for the FM interlayer coupling, which is positive for the FM interlayer coupling while negative for the AFM coupling. Here, $J_{\bot }$ only couples the sites of the first layer with the sites of the second layer that are positioned directly next to them. The findings presented in Ref. \cite{Hua2023PRB} also demonstrate the impact of incorporating spatially modulated remote interlayer couplings.

In the FM case, the classical ground state is represented by the uniform state $\mathbf{S}_{i,l}\equiv S \mathbf{\hat{z}}$, where the spins point along the $+z$ direction. In the ordered phase that emerges at sufficiently low temperatures, an effective magnon Hamiltonian using linear spin-wave theory was observed. By employing the Holstein-Primakoff transformation
$S_{i}^{+}=S^{x}_{i}+iS^{y}_{i}\simeq \sqrt{2S}d_{i}, S_{i}^{-}=S^{x}_{i}-iS^{y}_{i}\simeq \sqrt{2S} d_{i}^{\dag },  S_{i}^{z}=S-d_{i}^{\dag }d_{i}$, 
and neglecting magnon-magnon interactions, the spin Hamiltonian can be transformed into a quadratic magnon Hamiltonian
\begin{align}
\label{H}
H=&3JS \sum_{i,l}d_{i,l}^{\dag }d_{i,l}-JS\sum_{\left\langle i,j\right\rangle ,l}( d_{i,l}^{\dag}d_{j,l}+\text{H.c}.)  \\
&+J_{\bot}S \sum_{\left\langle i,j\right\rangle }[(d_{i,2}^{\dag }d_{i,2}+d_{j,1}^{\dag }d_{j,1})-(d_{i,2}^{\dag }d_{j,1}+\text{H.c.})], \nonumber
\end{align}
where $d_{i}^{\dag }$ ($d_{i}$) is the bosonic creation (annihilation) operator. Furthermore, through the Fourier transformation, a $28 \times 28$ magnon Hamiltonian in the $k$ space was derived as $H=\sum_{\mathbf{k}} \Psi _{\mathbf{k}}^{\dag } H_{\mathbf{k}} \Psi _{\mathbf{k}}$, where the basis is given by $\Psi _{\mathbf{k}}^{\dag }=( c_{\mathbf{k},1}^{\dag },\cdots ,c_{\mathbf{k},28}^{\dag })$. The numerical values in the basis align with the site indicators within the unit cell, as depicted in Fig. 2(b). The concrete expression of $H_{\mathbf{k}}$ and additional details can be found in Ref.~\cite{Hua2023PRB}.

\begin{center}
\includegraphics[width=10cm]{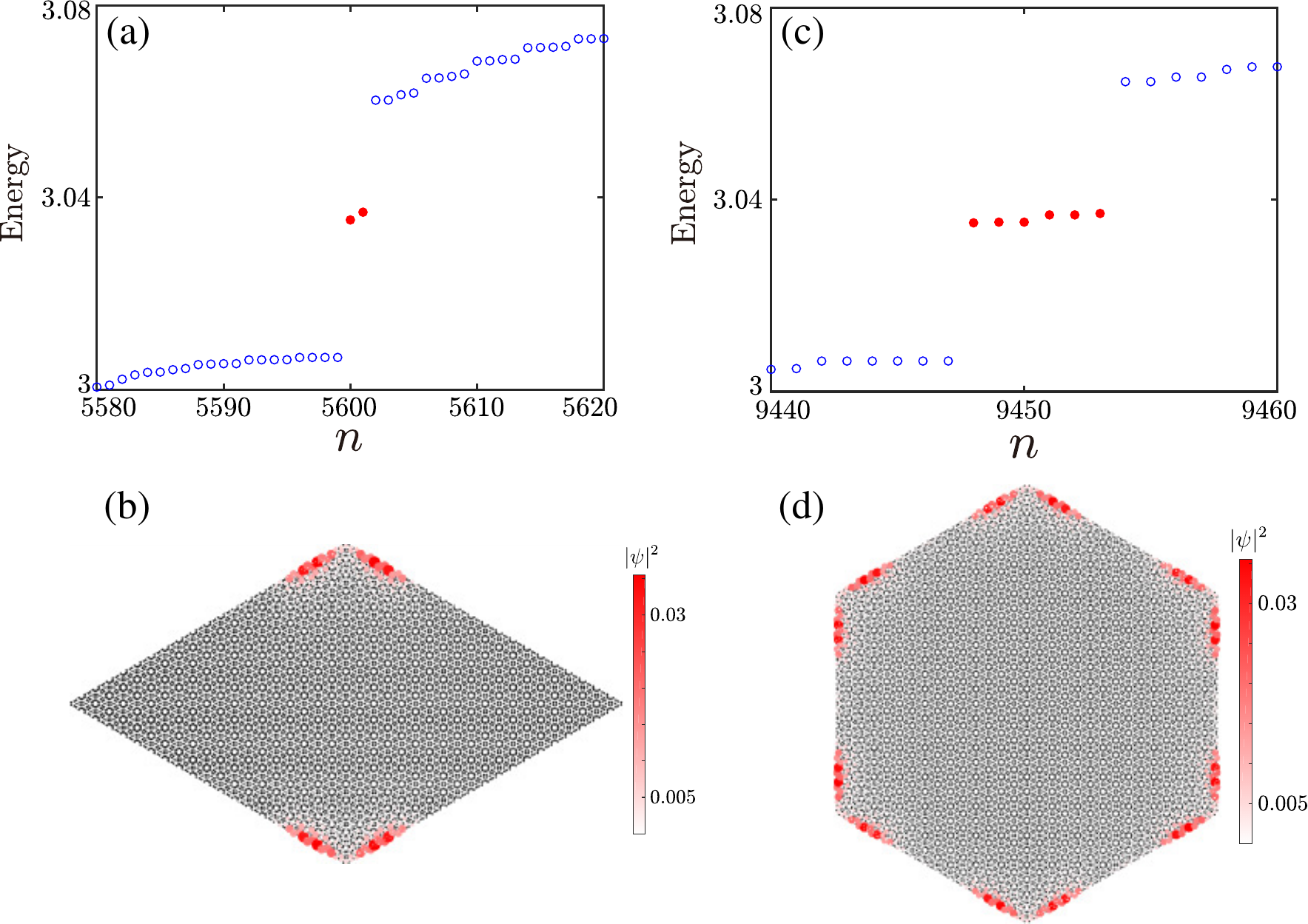}\\[5pt]  
\parbox[c]{15.0cm}{\footnotesize{\bf Fig.~5.} Higher-order topological magnon insulators of the TBHM with twist angle $21.78^\circ$\ucite{Hua2023PRB}. (a) Magnon energy spectrum of the Hamiltonian [Eq. (2)] for the rhombus-shaped TBHM system versus the eigenvalue index $n$. Red dots mark the in-gap magnon corner states. (b) The spatial distribution of the probability density of the two in-gap states. Magnon energy spectrum (c) and the spatial distribution of the probability density of the six in-gap states of the TBHM system with regular hexagonal boundary.}
\end{center}

By numerically diagonalizing the magnon Hamiltonian Eq. (2) in real space, the magnon energy spectrum was plotted in Fig. 5. It was observed that the magnon energy spectrum exhibits an energy gap, and notably, two in-gap states are present within this energy gap [as depicted in Fig. 5(a)]. As illustrated in Fig. 5(b), these two in-gap states are symmetrically localized at the top and bottom corners of the rhombus, respectively. The presence of twofold symmetric in-gap corner states is a characteristic feature of the SOTMI in the TBHM, and is associated with a mirror winding number. Furthermore, the presence of six magnon corner states is demonstrated in Figs 5(c) and (d) when considering a finite hexagon-shaped TBHM sample.

The magnon band structure of the TBHM system with a finite FM and AFM interlayer exchange interaction along a high symmetry line of the moir{\'e} Brillouin zone obtained is depicted in Fig. 6 by numerically diagonalizing $H_{\mathbf{k}}$.
It is observed that the finite FM interlayer coupling results in the opening of a sizable energy gap at the point K, as shown in Figs. 6(a) and (b). This energy gap exhibits topologically non-trivial characteristics, as evidenced by the emergence of magnon corner states within it under open boundary conditions. A mirror winding number was utilized as a topological invariant to characterize the higher-order magnon topology. In addition, the AFM magnon band structures of the TBHM system are depicted in Figs. 6(c) and (d). It is noteworthy that, unlike the FM case, the AFM magnon energy bands of the system remain gapless and exhibit stable linear dispersion, irrespective of the presence of AFM interlayer interactions.

\begin{center}
\includegraphics[width=10cm]{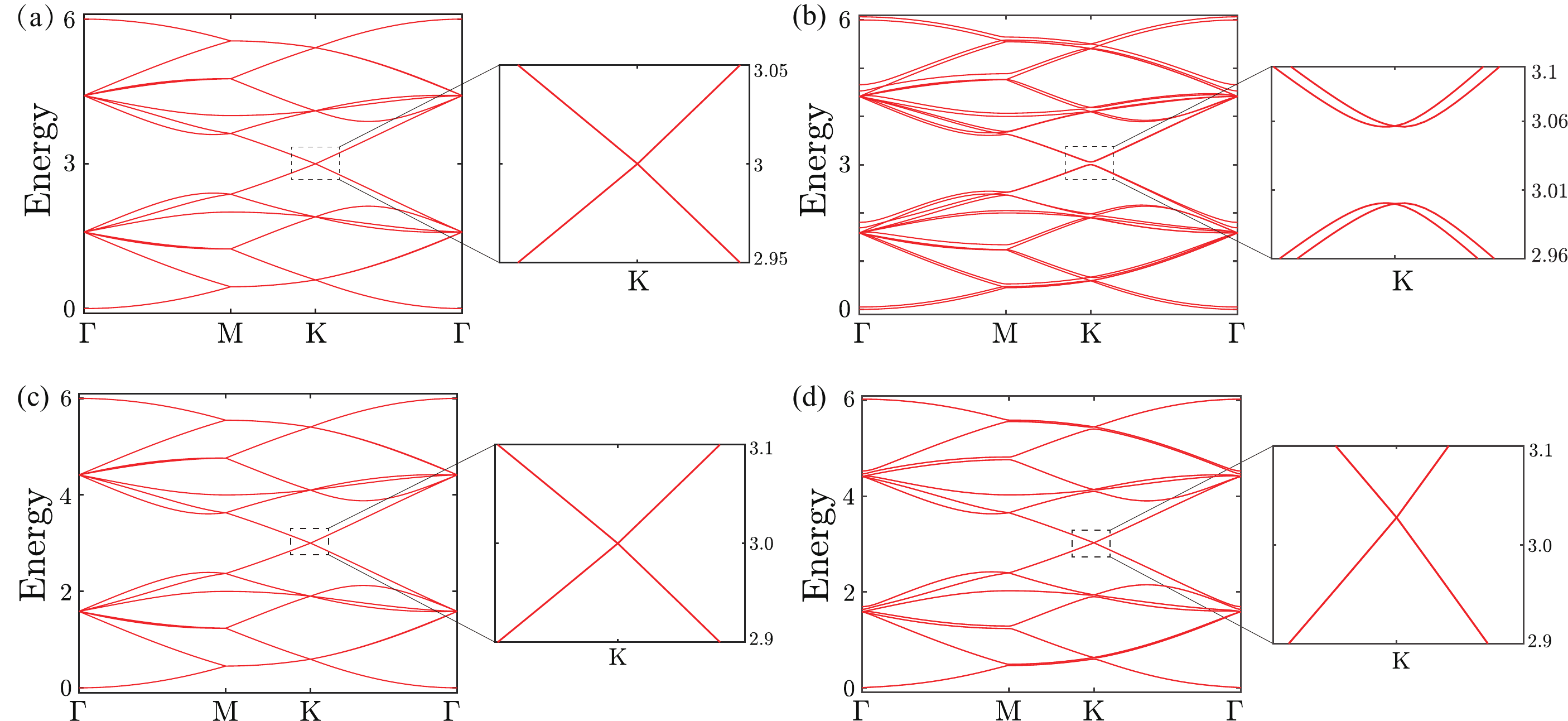}\\[5pt]  
\parbox[c]{15.0cm}{\footnotesize{\bf Fig.~6.} Higher-order topological magnon insulators of the TBHM with twist angle $21.78^\circ$\ucite{Hua2023PRB}. Magnon band structures for the FM interlayer interaction amplitude (a) $J_{\bot }/J=0$ and (b) $J_{\bot }/J=0.2$, as well as for the AFM interlayer interaction amplitude (c) $J_{\bot }/J=0$ and (d) $J_{\bot }/J=-0.2$. The zoom-ins demonstrate the dispersion around the K point.}
\end{center}

\section{Higher-order topology in various classical twisted systems}

We are currently poised to review HOTPs in various classical twisted systems, such as acoustic, photonic, and mechanical systems. In the realm of acoustic systems, the physical manifestation of acoustic topological corner states has been achieved through moir\'{e} twisting and ultrastrong interlayer couplings in bilayer honeycomb lattices composed of coupled acoustic cavities\ucite{Wu2022PRAp}. The ultrastrong interlayer couplings result in a significant acoustic band gap exhibiting a distinctive higher-order band topology, characterized by unprecedented topological indices and layer-hybridized corner states. The presence of higher-order topological edge and corner states is confirmed through acoustic pump-probe measurements. In photonic systems, a proposal has been made for a bilayer photonic graphene with even sublattice exchange symmetry to demonstrate a band gap suitable for the generation of topologically protected corner states\ucite{Oudich2021PRB}. Furthermore, through explicit manipulation of the underlying reflection symmetry at the boundary terminations of twisted bilayer photonic moir\'{e} superlattices, first-order topological edge modes, corresponding to nontrivial flat bands induced by strong interlayer coupling, naturally transition into higher-order topological corner modes\ucite{Yi2022Light}. In addition, the occurrence of corner modes in elastic twisted kagome lattices at a critical twist angle was also examined\ucite{Danawe2021PRB} systems.
 
\section{Conclusions and perspectives}

In this review, we have introduced the recent progress on HOTPs in 2D twisted van der Waals and twisted metamaterials. We begin by discuss various types of HOTPs in 2D electronic twisted van der Waals materials, encompassing higher-order topological insulator phases, higher-order topological superconductor phases, higher-order topological semimetal phases, and quasicrystalline higher-order topological insulator phases. Subsequently, we delve into a discussion on the SOTMI featuring magnon corner states in a TBHM with a large commensurate angle. Furthermore, we present the advancements of HOTPs on other classical systems including acoustic, photonic, and mechanical twisted systems.

Here, this review has focused solely on a restricted selection of 2D twisted systems for HOTPs, with numerous other aspects left unexplored within this discourse. And, the investigation of numerous unresolved issues remains a valuable pursuit in academic research. For instance, there is limited literature on the Floquet HOTPs with anomalous boundary states in 2D electronic twisted van der Waals materials. Furthermore, it is crucial to comprehend HOTPs with non-Hermitian, nonlinear, non-Abelian, and defects in these systems, surpassing the limitations of traditional topology.

\addcontentsline{toc}{chapter}{Acknowledgment}
\section*{Acknowledgment}
This research was supported the National Natural Science Foundation of China (Grant Nos. 12304539, 12074108, 12474151, and 12347101), and  the Natural Science Foundation of Chongqing (Grant No.~CSTB2022NSCQ-MSX0568).

\newpage


\addcontentsline{toc}{chapter}{References}
\begin{thebibliography}{99}\footnotesize
\itemsep=-3pt plus.2pt minus.2pt   

\bibitem{Benalcazar2017Science} Benalcazar W A, Bernevig B A and Hughes T L \href{https://www.science.org/doi/10.1126/science.aah6442}{2017 \emph{Science} \textbf{357} 61-66}
\bibitem{Benalcazar2017PRB} Benalcazar W A, Bernevig B A and Hughes T L \href{https://doi.org/10.1103/PhysRevB.96.245115}{2017 \emph{Phys. Rev. B} \textbf{96} 245115}
\bibitem{Langbehn2017PRL} Langbehn J, Peng Y, Trifunovic L, von Oppen F and Brouwer P W \href{https://doi.org/10.1103/PhysRevLett.119.246401}{2017 \emph{Phys. Rev. Lett.} \textbf{119} 246401}
\bibitem{Song2017PRL} Song Z, Fang Z and Fang C \href{https://doi.org/10.1103/PhysRevLett.119.246402}{2017 \emph{Phys. Rev. Lett.} \textbf{119} 246402}
\bibitem{Schindler2018SA} Schindler F, Cook A M, Vergniory M G, Wang Z, Parkin S S P, Bernevig B A and Neupert T \href{https://www.science.org/doi/10.1126/sciadv.aat0346}{2018 \emph{Sci. Adv.} \textbf{4} eaat0346}
\bibitem{Yan2019Wlxb} Yan Z B \href{https://wulixb.iphy.ac.cn/article/doi/10.7498/aps.68.20191101}{2019 \emph{Acta Phys. Sin.} \textbf{68} 226101}
\bibitem{Xie2021NRP} Xie B, Wang H X, Zhang X, Zhan P, Jiang J H, Lu M and Chen Y \href{https://doi.org/10.1038/s42254-021-00323-4}{2021 \emph{Nat. Rev. Phys.} \textbf{3} 520-532}
\bibitem{Yang2024JPCM} Yang Y B, Wang J H, Li K and Xu Y \href{https://doi.org/10.1088/1361-648X/ad3abd}{2024 \emph{J. Phys. Condens. Matter} \textbf{36} 283002}
\bibitem{Wieder2022NRM} Wieder B J, Bradlyn B, Cano J, Wang Z, Vergniory M G, Elcoro L, Soluyanov A A, Felser C, Neupert T, Regnault N and Bernevig B A \href{https://doi.org/10.1038/s41578-021-00380-2}{2022 \emph{Nat. Rev. Mater.} \textbf{7} 196-216}

\bibitem{Garcia2018Nature} Garcia M S, Peri V, Susstrunk R, Bilal O R, Larsen T, Villanueva L G and Huber S D \href{https://doi.org/10.1038/nature25156}{2018 \emph{Nature} \textbf{555} 342-345}
\bibitem{Peterson2018Nature} Peterson C W, Benalcazar W A, Hughes T L and Bahl G \href{https://doi.org/10.1038/nature25777}{2018 \emph{Nature} \textbf{555} 346-350}
\bibitem{Schindler2018NP} Schindler F, Wang Z, Vergniory M G, Cook A M, Murani A, Sengupta S, Kasumov A Y, Deblock R, S Jeon, Drozdov I, Bouchiat H, Gueron S, Yazdani A, Bernevig B A and Neupert T \href{https://doi.org/10.1038/s41567-018-0224-7}{2018 \emph{Nat. Phys.} \textbf{14} 918-924}
\bibitem{Imhof2018NP} Imhof S, Berger C, Bayer F, Brehm J, Molenkamp L W, Kiessling T, Schindler F, Lee C H, Greiter M, Neupert T and Thomale R \href{https://doi.org/10.1038/s41567-018-0246-1}{2018 \emph{Nat. Phys.} \textbf{14} 925-929}
\bibitem{Xu2019PRLHOTI} Xu Y , Song Z , Wang Z , Weng H and Dai X \href{https://doi.org/10.1103/PhysRevLett.122.256402}{2019 \emph{Phys. Rev. Lett.} \textbf{122} 256402}
\bibitem{Yue2019NP} Yue C, Xu Y, Song Z, Weng H, Lu Y M, Fang C and Dai X \href{https://doi.org/10.1038/s41567-019-0457-0}{2019 \emph{Nat. Phys.} \textbf{15} 577-581}
\bibitem{Xue2019NM} Xue H, Yang Y, Gao F, Chong Y and Zhang B \href{https://doi.org/10.1038/s41563-018-0251-x}{2019 \emph{Nat. Mater.} \textbf{18} 108-112}
\bibitem{Ni2019NM} Ni X, Weiner M, Alu A and Khanikaev A B \href{https://doi.org/10.1038/s41563-018-0252-9}{2019 \emph{Nat. Mater.} \textbf{18} 113-120}
\bibitem{Sheng2019PRL} Sheng X L, Chen C, Liu H, Chen Z, Yu Z M, Zhao Y X and Yang S A \href{https://doi.org/10.1103/PhysRevLett.123.256402}{2019 \emph{Phys. Rev. Lett.} \textbf{123} 256402}
\bibitem{Ren2020PRL} Ren Y, Qiao Z and Niu Q \href{https://doi.org/10.1103/PhysRevLett.124.166804}{2020 \emph{Phys. Rev. Lett.} \textbf{124} 166804}
\bibitem{Zhang2020PRL} Zhang R X, Wu F and Sarma S D \href{https://doi.org/10.1103/PhysRevLett.124.136407}{2020 \emph{Phys. Rev. Lett.} \textbf{124} 136407}
\bibitem{Chen2021PRL} Chen R, Liu T, Wang C M, Lu H Z and Xie X C \href{https://doi.org/10.1103/PhysRevLett.127.066801}{2021 \emph{Phys. Rev. Lett.} \textbf{127} 066801}
\bibitem{Ning2022PRB} Ning Z, Fu B, Xu D H and Wang R \href{https://doi.org/10.1103/PhysRevB.105.L201114}{2022 \emph{Phys. Rev. B} \textbf{105} L201114}
\bibitem{ChenR2023PRB} Chen R , Zhou B and Xu D H \href{https://doi.org/10.1103/PhysRevB.108.195306}{2023 \emph{Phys. Rev. B} \textbf{108} 195306}
\bibitem{XuXX2024CPB} Xu X X, Wang Z M, Xu D H and Chen C Z \href{https://iopscience.iop.org/article/10.1088/1674-1056/ad4634/meta}{2024 \emph{Chin. Phys. B} \textbf{33} 067801}
\bibitem{Zhan2024NL} Zhan F, Qin Z, Xu D H, Zhou X, Ma D S, Wang R \href{https://doi.org/10.1021/acs.nanolett.4c01817}{2024 \emph{Nano Lett.} \textbf{24} 7741-7747}
\bibitem{ChneH2023CPB} Chen H, Liu Z R, Chen R and Zhou B \href{https://iopscience.iop.org/article/10.1088/1674-1056/ad09d4/meta}{2023 \emph{Chin. Phys. B} \textbf{33} 017202}
\bibitem{YanZB2018PRL} Yan Z, Song F and Wang Z \href{https://doi.org/10.1103/PhysRevLett.121.096803}{2018 \emph{Phys. Rev. Lett.} \textbf{121} 096803}
\bibitem{YanZB2019PRL} Yan Z \href{https://doi.org/10.1103/PhysRevLett.123.177001}{2019 \emph{Phys. Rev. Lett.} \textbf{123} 177001}
\bibitem{YanZB2019PRB} Yan Z \href{https://doi.org/10.1103/PhysRevB.100.205406}{2019 \emph{Phys. Rev. B} \textbf{100} 205406}
\bibitem{ZhuD2023PRB} Zhu D, Kheirkhah M and Yan Z \href{https://doi.org/10.1103/PhysRevB.107.085407}{2023 \emph{Phys. Rev. B} \textbf{107} 085407}
\bibitem{ZengQB2020PRB} Zeng Q B, Yang Y B and Xu Y \href{https://doi.org/10.1103/PhysRevB.101.241104}{2020 \emph{Phys. Rev. B} \textbf{101} 241104(R)}
\bibitem{YangYB2020PRR} Yang Y B, Li K, Duan L M and Xu Y \href{https://doi.org/10.1103/PhysRevResearch.2.033029}{2020 \emph{Phys. Rev. Research} \textbf{2} 033029}
\bibitem{Tao2020NJP} Tao Y L, Dai N, Yang Y B, Zeng Q B and Xu Y \href{https://iopscience.iop.org/article/10.1088/1367-2630/abc1f9}{2020 \emph{New J. Phys.} \textbf{22} 103058}
\bibitem{YangYB2021PRB} Yang Y B, Li K, Duan L M and Xu Y \href{https://doi.org/10.1103/PhysRevB.103.085408}{2021 \emph{Phys. Rev. B} \textbf{103} 085408}
\bibitem{PengY2022OL} Peng Y, Liu E, Yan B, Xie J, Shi A, Peng P, Li H and Liu J \href{https://doi.org/10.1364/OL.457058}{2022 \emph{Opt. Lett.} \textbf{47} 3011-3014}
\bibitem{ShiA2024PRB} Shi A, Peng Y, Peng P, Chen J and Liu J \href{https://doi.org/10.1103/PhysRevB.110.014106}{2024 \emph{Phys. Rev. B} \textbf{110} 014106}
\bibitem{ZhaoPL2021PRL} Zhao P L, Qiang X B, Lu H Z and Xie X C \href{https://doi.org/10.1103/PhysRevLett.127.176601}{2021 \emph{Phys. Rev. Lett.} \textbf{127} 176601}
\bibitem{LiuT2018PRB} Liu T, He J J and Nori F \href{https://doi.org/10.1103/PhysRevB.98.245413}{2018 \emph{Phys. Rev. B} \textbf{98} 245413}
\bibitem{LiuT2019PRL} Liu T, Zhang Y R, Ai Q, Gong Z, Kawabata K, Ueda M and Nori F \href{https://doi.org/10.1103/PhysRevLett.122.076801}{2019 \emph{Phys. Rev. Lett.} \textbf{122} 076801}
\bibitem{LiuT2021PRL} Liu T, He J J, Yang Z and Nori F \href{https://doi.org/10.1103/PhysRevLett.127.196801}{2021 \emph{Phys. Rev. Lett.} \textbf{127} 196801}
\bibitem{LuC2023PRB} Lu C, Cai Z F, Zhang M, Wang H, Ai Q and Liu T \href{https://doi.org/10.1103/PhysRevB.107.165403}{2023 \emph{Phys. Rev. B} \textbf{107} 165403}
\bibitem{QiY2020PRL} Qi Y, Qiu C, Xiao M, He H, Ke M and Liu Z \href{https://doi.org/10.1103/PhysRevLett.124.206601}{2020 \emph{Phys. Rev. Lett.} \textbf{124} 206601}
\bibitem{WuJ2020PRB} Wu J, Huang X, Lu J, Wu Y, Deng W, Li F and Liu Z \href{https://doi.org/10.1103/PhysRevB.102.104109}{2020 \emph{Phys. Rev. B} \textbf{102} 104109}
\bibitem{YangY2021PRL} Yang Y, Lu J, Yan M, Huang X, Deng W and Liu Z \href{https://doi.org/10.1103/PhysRevLett.126.156801}{2021 \emph{Phys. Rev. Lett.} \textbf{126} 156801}
\bibitem{WeiQ2021PRL} Wei Q, Zhang X, Deng W, Lu J, Huang X, Yan M, Chen G, Liu Z and Jia S \href{https://doi.org/10.1103/PhysRevLett.127.255501}{2021 \emph{Phys. Rev. Lett.} \textbf{127} 255501}
\bibitem{WuJ2022PRB} Wu J, Huang X, Yang Y, Deng W, Lu J, Deng W and Liu Z \href{https://doi.org/10.1103/PhysRevB.105.195127}{2022 \emph{Phys. Rev. B} \textbf{105} 195127}
\bibitem{WangZ2024CP} Wang Z, Ye L, Pu Z, Ma Q, He H, Lu J, Deng W, Huang X, Ke M and Liu Z \href{https://doi.org/10.1038/s42005-024-01681-y}{2024 \emph{Commun. Phys.} \textbf{7} 193}
\bibitem{YeL2024PRL} Ye L, Wang Q, Fu Z, He H, Huang X, Ke M, Lu J, Deng W and Liu Z \href{https://doi.org/10.1103/PhysRevLett.133.126602}{2024 \emph{Phys. Rev. Lett.} \textbf{133} 126602}
\bibitem{HuangX2022SB} Huang X, Lu J, Yan Z, Yan M, Deng W, Chen G and Liu Z  \href{https://doi.org/10.1016/j.scib.2021.11.020}{2022 \emph{Sci. Bull.} \textbf{5} 488-494}
\bibitem{QiY2022APL} Qi Y, He H and Xiao M \href{https://doi.org/10.1063/5.0095543}{2022 \emph{Appl. Phys. Lett.} \textbf{120} 212202}
\bibitem{QiY2024PRB} Qi Y, He Z, Deng K, Li J and Wang Y \href{https://doi.org/10.1103/PhysRevB.109.L060101}{2024 \emph{Phys. Rev. B} \textbf{109} L060101}
\bibitem{PuZ2023PRL} Pu Z, He H, Luo L, Ma Q, Ye L, Ke M and Liu Z \href{https://doi.org/10.1103/PhysRevLett.130.116103}{2023 \emph{Phys. Rev. Lett.} \textbf{130} 116103}
\bibitem{MaQ2024PRL} Ma Q, Pu Z, Ye L, Lu J, Huang X, Ke M, He H, Deng W and Liu Z \href{https://doi.org/10.1103/PhysRevLett.132.066601}{2024 \emph{Phys. Rev. Lett.} \textbf{132} 066601}
\bibitem{LuoXJ2023PRB} Luo X J, Pan X H, Liu C X and Liu X \href{https://doi.org/10.1103/PhysRevB.107.045118}{2023 \emph{Phys. Rev. B} \textbf{107} 045118}
\bibitem{LiuF2023PRB} Liu F \href{https://doi.org/10.1103/PhysRevB.108.245140}{2023 \emph{Phys. Rev. B} \textbf{108} 245140}
\bibitem{LuoXJ2023PRB2} Luo X J and Wu F \href{https://doi.org/10.1103/PhysRevB.108.075143}{2023 \emph{Phys. Rev. B} \textbf{108} 075143}
\bibitem{HuYS2021PRB} Hu Y S, Ding Y R, Zhang J, Zhang Z Q and Chen C Z \href{https://doi.org/10.1103/PhysRevB.104.094201}{2021 \emph{Phys. Rev. B} \textbf{104} 094201}
\bibitem{LiuZ2022PRB} Liu Z, Ren Y, Han Y, Niu Q and Qiao Z \href{https://doi.org/10.1103/PhysRevB.106.195303}{2022 \emph{Phys. Rev. B} \textbf{106} 195303}
\bibitem{LiuZR2021PRB} Liu Z R, Hu L H, Chen C Z, Zhou B and Xu D H \href{https://doi.org/10.1103/PhysRevB.103.L201115}{2021 \emph{Phys. Rev. B} \textbf{103} L201115}
\bibitem{LuC2023PRB2} Lu C, Zhang M, Wang H, Ai Q and Liu T \href{https://doi.org/10.1103/PhysRevB.107.125118}{2023 \emph{Phys. Rev. B} \textbf{107} 125118}
\bibitem{HuY2024PRB} Hu Y, Liu S, Pan B, Zhou P and Sun L \href{https://doi.org/10.1103/PhysRevB.109.L121403}{2024 \emph{Phys. Rev. B} \textbf{109} L121403}
\bibitem{ChenR2024PRB} Chen R, Zhou B and Xu D H \href{https://doi.org/10.1103/PhysRevB.110.L121301}{2024 \emph{Phys. Rev. B} \textbf{110} L121301}
\bibitem{PengT2024PRB} Peng T, Xiong Y C, Hua C B, Liu Z R, Zhu X, Cao W, Lv F, Hou Y, Zhou B, Wang Z and Xiong R \href{https://doi.org/10.1103/PhysRevB.109.195301}{2024 \emph{Phys. Rev. B} \textbf{109} 195301}
\bibitem{WuYJ2022PRA} Wu Y J, He W, Li N, Li Z and Hou J \href{https://doi.org/10.1103/PhysRevA.106.063524}{2022 \emph{Phys. Rev. A} \textbf{106} 063524}
\bibitem{LiCA2022FP} Li C A \href{https://doi.org/10.3389/fphy.2022.861242}{2022 \emph{Front. Phys.} \textbf{10} 861242}
\bibitem{ChneXT2023CPB} Chen X T, Liu C H, Xu D H and Chen C Z \href{https://iopscience.iop.org/article/10.1088/0256-307X/40/9/097403/meta}{2023 \emph{Chin. Phys. Lett.} \textbf{40} 097403}
\bibitem{LiJ2023APL} Li J, Kuang M, Bai J, Ding G, Yuan H, Xie C, Wang W and Wang X \href{https://doi.org/10.1063/5.0158822}{2023 \emph{Appl. Phys. Lett.} \textbf{123} 012201}
\bibitem{Guo2023MTP} Guo Z, Liu Y, Jiang H, Zhang X, Jin L, Liu C and Liu G \href{https://doi.org/10.1016/j.mtphys.2023.101153}{2023 \emph{Materials Today Physics} \textbf{36} 101153}
\bibitem{Miao2023PRB} Miao C M, Wan Y H, Sun Q F and Zhang Y T \href{https://doi.org/10.1103/PhysRevB.108.075401}{2023 \emph{Phys. Rev. B} \textbf{108} 075401}
\bibitem{Scammell2022PRB} Scammell H D, Ingham J, Geier M and Li T \href{https://doi.org/10.1103/PhysRevB.105.195149}{2022 \emph{Phys. Rev. B} \textbf{105} 195149}
\bibitem{Habel2024PRB} Habel J, Mook A, Willsher J and Knolle J \href{https://doi.org/10.1103/PhysRevB.109.024441}{2024 \emph{Phys. Rev. B} \textbf{109} 024441}


\bibitem{Varjas2019PRL} Varjas D, Lau A, Poyhonen K, Akhmerov A R, Pikulin D I and Fulga I C \href{https://doi.org/10.1103/PhysRevLett.123.196401}{2019 \emph{Phys. Rev. Lett.} \textbf{123} 196401}
\bibitem{Chen2020PRL} Chen R, Chen C Z, Gao J H, Zhou B and Xu D H \href{https://doi.org/10.1103/PhysRevLett.124.036803}{2020 \emph{Phys. Rev. Lett.} \textbf{124} 036803}
\bibitem{Hua2020PRB} Hua C B, Chen R, Zhou B and Xu D H \href{https://doi.org/10.1103/PhysRevB.102.241102}{2020 \emph{Phys. Rev. B} \textbf{102} 241102(R)}
\bibitem{Huang2021NL} Huang H, Fan J, Li D and Liu F \href{https://doi.org/10.1021/acs.nanolett.1c02661}{2021 \emph{Nano Lett.} \textbf{21} 7056-62}
\bibitem{Peng2021PRB} Peng T, Hua C B, Chen R, Liu Z R, Xu D H and Zhou B \href{https://doi.org/10.1103/PhysRevB.104.245302}{2021 \emph{Phys. Rev. B} \textbf{104} 245302}
\bibitem{Wang2022PRL} Wang C, Liu F and Huang H \href{https://doi.org/10.1103/PhysRevLett.129.056403}{2022 \emph{Phys. Rev. Lett.} \textbf{129} 056403}
\bibitem{Xiong2022PRAp} Xiong L, Zhang Y, Liu Y, Zheng Y and Jiang X \href{https://doi.org/10.1103/PhysRevApplied.18.064089}{2022 \emph{Phys. Rev. Appl.} \textbf{18} 064089}
\bibitem{Mao2024PRB} Mao Y F , Tao Y L , Wang J H , Zeng Q B and Xu Y \href{https://doi.org/10.1103/PhysRevB.109.134205}{2024 \emph{Phys. Rev. B} \textbf{109} 134205}
\bibitem{Ouyang2024PRB} Ouyang C , He Q , Xu D H and Liu F \href{https://doi.org/10.1103/PhysRevB.110.075425}{2024 \emph{Phys. Rev. B} \textbf{110} 075425}
\bibitem{Lv2021CP} Lv B , Chen R , Li R , Guan C , Zhou B , Dong G , Zhao C , Li Y , Wang Y , Tao H , Shi J and Xu D H \href{https://doi.org/10.1038/s42005-021-00610-7}{2021 \emph{Commun. Phys.} \textbf{4} 108}
\bibitem{Spurrier2020PRR} Spurrier S and Cooper N R \href{https://doi.org/10.1103/PhysRevResearch.2.033071}{2020 \emph{Phys. Rev. Research} \textbf{2} 033071}
\bibitem{Traverso2022Symmetry} Traverso S, Traverso Ziani N and Sassetti M \href{https://doi.org/10.3390/sym14081736}{2022 \emph{Symmetry} \textbf{14} 1736}
\bibitem{Traverso2022PRB} Traverso S, Sassetti M and Ziani N T \href{https://doi.org/10.1103/PhysRevB.106.125428}{2022 \emph{Phys. Rev. B} \textbf{106} 125428}
\bibitem{Shi2024Laser} Shi A, Peng Y, Jiang J, Peng Y, Peng P, Chen J, Chen H, Wen S, Lin X, Gao F and Liu J \href{https://doi.org/10.1002/lpor.202300956}{2024 \emph{Laser Photonics Rev.} \textbf{18} 2300956}
\bibitem{Agarwala2020PRR} Agarwala A , Juricic V and Roy B \href{https://doi.org/10.1103/PhysRevResearch.2.012067}{2020 \emph{Phys. Rev. Research} \textbf{2} 012067(R)}
\bibitem{Wang2021PRLAmorphous} Wang J H , Yang Y B , Dai N and Xu Y \href{https://doi.org/10.1103/PhysRevLett.126.206404}{2021 \emph{Phys. Rev. Lett.} \textbf{126} 206404}
\bibitem{Peng2022PRB} Peng T , Hua C B , Chen R , Liu Z R, Huang H M and Zhou B \href{https://doi.org/10.1103/PhysRevB.106.125310}{2022 \emph{Phys. Rev. B} \textbf{106} 125310}
\bibitem{Tao2023Post} Tao Y L, Wang J H and Xu Y \href{https://scipost.org/10.21468/SciPostPhys.15.5.193}{2023 \emph{SciPost Phys.} \textbf{15} 193}
\bibitem{Manna2022PRB} Manna S, Nandy S and Roy B \href{https://doi.org/10.1103/PhysRevB.105.L201301}{2022 \emph{Phys. Rev. B} \textbf{105} L201301}
\bibitem{Zheng2022SB} Zheng S, Man X, Kong Z-L, Lin Z-K, Duan G, Chen N, Yu D, Jiang J-H and Xia B \href{https://doi.org/10.1016/j.scib.2022.09.020}{2022 \emph{Sci. Bull.} \textbf{67} 2069-75}
\bibitem{ChenH2023CPB} Chen H, Liu Z-R, Chen R and Zhou B \href{https://iopscience.iop.org/article/10.1088/1674-1056/ad09d4}{2023 \emph{Chin. Phys. B} \textbf{33} 017202}
\bibitem{Tao2023PRB} Tao Y L and Xu Y \href{https://doi.org/10.1103/PhysRevB.107.184201}{2023 \emph{Phys. Rev. B} \textbf{107} 184201}
\bibitem{Liu2023PRB} Liu Z R, Hua C B, Peng T, Chen R and Zhou B \href{https://doi.org/10.1103/PhysRevB.107.125302}{2023 \emph{Phys. Rev. B} \textbf{107} 125302}


\bibitem{Cao2018Nature1} Cao Y, Fatemi V, Fang S, Watanabe K, Taniguchi T, Kaxiras E and Jarillo-Herrero P \href{https://doi.org/10.1038/nature26160}{2018 \emph{Nature} \textbf{556} 43}
\bibitem{Cao2018Nature2} Cao Y, Fatemi V, Demir A, Fang S, Tomarken S L, Luo J Y, Sanchez-Yamagishi J D, Watanabe K, Taniguchi T, Kaxiras E, Ashoori R C and Jarillo-Herrero P \href{https://doi.org/10.1038/nature26154}{2018 \emph{Nature} \textbf{556} 80}
\bibitem{Bistritzer2011PNAS} Bistritzer R and MacDonald A H \href{https://doi.org/10.1073/pnas.1108174108}{2011 \emph{Proc. Natl. Acad. Sci. USA} \textbf{108} 12233}
\bibitem{Serlin2020Science} Serlin M, Tschirhart C L, Polshyn H, Zhang Y, Zhu J, Watanabe K, Taniguchi T, Balents L, Young A F \href{https://www.science.org/doi/10.1126/science.aay5533}{2020 \emph{Science} \textbf{367} 900-903}
\bibitem{Wu2019PRL} Wu F, Lovorn T, Tutuc E, Martin I and MacDonald A H \href{https://doi.org/10.1103/PhysRevLett.122.086402}{2019 \emph{Phys. Rev. Lett.} \textbf{122} 086402}


\bibitem{Park2019PRL} Park M J, Kim Y, Cho G Y and Lee S \href{https://doi.org/10.1103/PhysRevLett.123.216803}{2019 \emph{Phys. Rev. Lett.} \textbf{123} 216803}
\bibitem{Liu2021PRL} Liu B, Xian L, Mu H, Zhao G, Liu Z, Rubio A and Wang Z F \href{https://doi.org/10.1103/PhysRevLett.126.066401}{2021 \emph{Phys. Rev. Lett.} \textbf{126} 066401}
\bibitem{Park2021Carbon} Park M J, Jeon S, Lee S, Park H C and Kim Y \href{https://doi.org/10.1016/j.carbon.2020.12.037}{2021 \emph{Carbon} \textbf{175} 260-265}
\bibitem{Liu2022PRB} Liu B B, Zeng X T, C Chen, Chen Z and Sheng X L \href{https://doi.org/10.1103/PhysRevB.106.035153}{2022 \emph{Phys. Rev. B} \textbf{106} 035153}
\bibitem{Lim2023njpCM} Kim S W, Jeon S, Park M J and Kim Y \href{https://doi.org/10.1038/s41524-023-01105-5}{2023 \emph{npj Comput. Mater.} \textbf{9} 152}
\bibitem{Chew2023PRB} Chew A, Wang Y, Bernevig B A and Song Z D \href{https://doi.org/10.1103/PhysRevB.107.094512}{2023 \emph{Phys. Rev. B} \textbf{107} 094512}
\bibitem{Qian2023PRB} Qian S, Li Y and Liu C C \href{https://doi.org/10.1103/PhysRevB.108.L241406}{2023 \emph{Phys. Rev. B} \textbf{108} L241406}
\bibitem{Hua2023PRB} Hua C B, Xiao F, Liu Z R, Sun J H, Gao J H, Chen C Z, Tong Q, Zhou B and Xu D H \href{https://doi.org/10.1103/PhysRevB.107.L020404}{2023 \emph{Phys. Rev. B} \textbf{107} L020404}
\bibitem{Wu2022PRAp} Wu S Q, Lin Z K, Jiang B, Zhou X, Hang Z H, Hou B and Jiang J H \href{https://doi.org/10.1103/PhysRevApplied.17.034061}{2022 \emph{Phys. Rev. Applied} \textbf{17} 034061}
\bibitem{Oudich2021PRB} Oudich M, Su G, Deng Y, Benalcazar W, Huang R, Gerard N J R K, Lu M, Zhan P and Jing Y \href{https://doi.org/10.1103/PhysRevB.103.214311}{2021 \emph{Phys. Rev. B} \textbf{103} 214311}
\bibitem{Yi2022Light} Yi C H, Park H C and Park M J \href{https://doi.org/10.1038/s41377-022-00977-4}{2022 \emph{Light Sci. Appl.} \textbf{11} 289}
\bibitem{Danawe2021PRB} Danawe H, Li H, Ba'ba'a H A and Tol S \href{https://doi.org/10.1103/PhysRevB.104.L241107}{2021 \emph{Phys. Rev. B} \textbf{104} L241107}

\bibitem{Ahn2018Science} Ahn S J, Moon P, Kim T H, Kim H W, Shin H C, Kim E H, Cha H W, Kahng S J, Kim P, Koshino M, Son Y W, Yang C W and Ahn J R \href{https://www.science.org/doi/10.1126/science.aar8412}{2018 \emph{Science} \textbf{361} 782}
\bibitem{Yao2018PNAS} Yao W, Wang E, Bao C, Zhang Y, Zhang K, Bao K, Chan C K, Chen C, Avila J, Asensio M C, Zhu J and Zhou S \href{https://www.pnas.org/doi/full/10.1073/pnas.1720865115}{2018 \emph{Proc. Natl. Acad. Sci. USA} \textbf{115} 6928}
\bibitem{Li2024Nature} Li Y, Zhang F, Ha V A, Lin Y C, Dong C, Gao Q, Liu Z, Liu X, Ryu S H, Kim H, Jozwiak C, Bostwick A, Watanabe K, Taniguchi T, Kousa B, Li X, Rotenberg E, Khalaf E, Robinson J A, Giustino F and Shih C K 
\href{https://doi.org/10.1038/s41586-023-06904-w}{2024 \emph{Nature} \textbf{625} 494-499}
\bibitem{Tsang2024Science} Tsang C S, Zheng X, Yang T, Yan Z, Han W, Wong L W, Liu H, Gao S, Leung K H, Lee C S, Lau S P, Yang M, Zhao J and Ly T H \href{https://www.science.org/doi/10.1126/science.adp7099}{2024 \emph{Science} \textbf{386} 198-205}

\bibitem{Sil2020JPCM} Sil A and Ghosh A K \href{https://iopscience.iop.org/article/10.1088/1361-648X/ab6f8b}{2020 \emph{J. Phys.: Condens. Matter} \textbf{32} 205601}
\bibitem{Hirosawa2020PRL} Hirosawa T, Diaz S A, Klinovaja J and Loss D \href{https://doi.org/10.1103/PhysRevLett.125.207204}{2020 \emph{Phys. Rev. Lett.} \textbf{125} 207204}
\bibitem{Mook2021PRB} Mook A, Diaz S A, linovaja J K and Loss D \href{https://doi.org/10.1103/PhysRevB.104.024406}{2021 \emph{Phys. Rev. B} \textbf{104} 024406}
\bibitem{Li2020PRB} Li Z X, Cao Y, Wang X R and Yan P \href{https://doi.org/10.1103/PhysRevB.101.184404}{2020 \emph{Phys. Rev. B} \textbf{101} 184404}
\bibitem{Li2021PRB} Li Z X, Wang Z, Zhang Z, Cao Y and Yan P \href{https://doi.org/10.1103/PhysRevB.103.214442}{2021 \emph{Phys. Rev. B} \textbf{103} 214442}
\bibitem{Park2021PRB} Park M J, Lee S and Kim Y B \href{https://doi.org/10.1103/PhysRevB.104.L060401}{2021 \emph{Phys. Rev. B} \textbf{104} L060401}
\bibitem{Li2022PRB} Li Y M, Wu Y J, Luo X W, Huang Y and Chang K \href{https://doi.org/10.1103/PhysRevB.106.054403}{2022 \emph{Phys. Rev. B} \textbf{106} 054403}
\bibitem{Bhowmik2024PRB} Bhowmik S , Banerjee S and Saha A \href{https://doi.org/10.1103/PhysRevB.109.104417}{2024 \emph{Phys. Rev. B} \textbf{109} 104417}
%
\bibitem{Ahn2024PQE} Ahn Y, Guo X, Son S, Sun Z, Zhao L \href{https://doi.org/10.1016/j.pquantelec.2024.100498}{2024 \emph{Progress in Quantum Electronics} \textbf{93} 100498}
\bibitem{ChengR2020PRB} Li Y H and Cheng R \href{https://doi.org/10.1103/PhysRevB.102.094404}{2020 \emph{Phys. Rev. B} \textbf{102} 094404}
\bibitem{Wang2020PRL} Wang C , Gao Y , Lv H , Xu X and Xiao D \href{https://doi.org/10.1103/PhysRevLett.125.247201}{2020 \emph{Phys. Rev. Lett.} \textbf{125} 247201}
\bibitem{Wang2023PRX} Wang H, Madami M, Chen J, Jia H, Zhang Y, Yuan R, Wang Y, He W, Sheng L, Zhang Y, Wang J, Liu S, Shen K, Yu G, Han X, Yu D, Ansermet J P, Gubbiotti G and Yu H \href{https://doi.org/10.1103/PhysRevX.13.021016}{2023 \emph{Phys. Rev. X} \textbf{13} 021016}
\bibitem{Bostrom2020Post} Bostrom E V, Claassen M, McIver J W, Jotzu G, Rubio A, Sentef M A \href{https://www.scipost.org/10.21468/SciPostPhys.9.4.061}{2020 \emph{SciPost Phys.} \textbf{9} 061}
\bibitem{Kim2023NL} Kim K M, Kiem D H, Bednik G, Han M J , Park M J \href{https://doi.org/10.1021/acs.nanolett.3c01529}{2023 \emph{Nano Lett.} \textbf{23} 6088-6094}
\bibitem{LiuJ2020PRB} Liu J, Wang L and K Shen \href{https://doi.org/10.1103/PhysRevB.102.144416}{2020 \emph{Phys. Rev. B} \textbf{102} 144416}
\bibitem{Zhu2021CPB} Zhu X, Guo H and S Feng \href{http://dx.doi.org/10.1088/1674-1056/abeee5}{2021 \emph{Chin. Phys. B} \textbf{30} 077505}
\bibitem{Ghader2020SR} Ghader D \href{https://doi.org/10.1038/s41598-020-72000-y}{2020 \emph{Sci. Rep.} \textbf{10} 15069}
\bibitem{Hejazi2020PNAS} Hejazi K, Luo Z X and Balents L \href{https://www.pnas.org/doi/full/10.1073/pnas.2000347117}{ \emph{Proc. Natl. Acad. Sci. USA} \textbf{117} 10721-10726}

\bibitem{Yin2022Nature}  Yin J X, Lian B and Hasan M Z \href{https://doi.org/10.1038/s41586-022-05516-0}{2022 \emph{Nature} \textbf{612} 647-657}
\bibitem{Neupert2022NP} Neupert T, Denner M M, Yin J X, Thomale R and Hasan M Z \href{https://doi.org/10.1038/s42254-023-00635-7}{ 2023 \emph{Nat. Rev. Phys.} \textbf{5} 635-658 }
\bibitem{Wang2023NRP}  Wang Y, Wu H, McCandless G T, Chan J Y and Ali M \href{https://doi.org/10.1038/s42254-023-00635-7}{ 2023 \emph{Nat. Rev. Phys.} \textbf{5} 635-658 }
\bibitem{Wilson2024NRM}  Wilson S D, Ortiz B R \href{https://doi.org/10.1038/s41578-024-00677-y}{ 2024 \emph{Nat. Rev. Mater.} \textbf{9} 420-432 }
\bibitem{Crasto2019PRB}  Crasto de Lima F, Miwa R H, and Suarez Morell E \href{https://link.aps.org/doi/10.1103/PhysRevB.100.155421}{ 2019 \emph{Phys. Rev. B} \textbf{100} 155421}
\bibitem{Wan2024arxiv}  Wan X, Zeng J, Zhu R, Xu D H, Zheng B and Wang R \href{https://arxiv.org/abs/2410.08487}{\emph{arXiv:2410.08487}}
\end{thebibliography}
\end{document}